\documentstyle[11pt,aaspp4]{article}

\newcommand{\kstar}{$\zeta^2$ CrB}

\begin{document}

\title{Zeta$^2$ Coronae Borealis, a Spectroscopic Triple System \\
Including an Asynchronous Close Binary}

\author{Karl D. Gordon and Christopher L. Mulliss}
\affil{Ritter Astrophysical Research Center, The University of Toledo
   Toledo, OH 43606 \\
Electronic mail: karlg@physics.utoledo.edu, cmulliss@physics.utoledo.edu}

\lefthead{Gordon \& Mulliss}
\righthead{Orbit for \kstar{}}

\begin{abstract}
Using the 1-m telescope at Ritter Observatory, we took 36 observations
of $\zeta^2$ Coronae Borealis with a fiber-fed \'echelle spectrograph.
From these observations, \kstar{} was found to be a triple system and
a new spectroscopic orbit was calculated.  This orbit has two periods,
a 1.72357 day period for the inner binary composed of \kstar{} A \& B
and a 251 day period for the outer binary composed of \kstar{} AB \&
C.  The inner binary is a double-lined spectroscopic binary composed
of two B7 V stars.  The inner binary's center of mass (\kstar{} AB)
describes a long-period single-lined variation identified with the
outer binary.  The inner binary period is significantly shorter than
the 12.5842 day period previously calculated by Abhyankar \& Sarma
(1966).  The inner binary possesses an essentially circular orbit ($e
= 0.01$) while the outer binary has an eccentric orbit ($e = 0.48$).
From the widths of their Si II 6371 \AA\ lines, the $v\sin i$'s were
calculated to be $46 \pm 7$ km s$^{-1}$ for \kstar{} A and $7.5 \pm 2$
km s$^{-1}$ for \kstar{} B.  As \kstar{} A \& B have similar masses,
their different rotational velocities make this system a sensitive
test of synchronization theories.
\end{abstract}


\section{Introduction}

  The B7 V star $\zeta^2$ Coronae Borealis (HD 139892, HR 5834, ADS
9737 A; $\alpha(2000) = 15^{\rm h}39^{\rm m}22\fs 66$, $\delta(2000) =
+36\arcdeg 38\arcmin 9\farcs 26$) has been known to be a double-lined
spectroscopic binary since the work of \cite{pla25}.  \kstar{} is the
brighter component ($\Delta m = 1.0$) of the visual double star ADS
9737 with a separation of $6\farcs 4$ from the fainter component
$\zeta^1$ CrB (\cite{hof82}).  Abhyankar \& Sarma (1966), hereafter
\cite{abh66}, refining the orbit of \cite{pla25} found a period of
12.5842 days from 58 observations.  They comment on the difficulty of
measuring the double broad lines of \kstar{}, which is evident in
their observed radial velocities' large deviations from their
calculated orbit (see Fig.\ 1 \& 2 of \cite{abh66}).  As a result,
their orbit received a `d' (poor orbit) classification in the {\it 8th
Catalogue of the Orbital Elements of Spectroscopic Binary Systems}
(\cite{bat89}).

  In an effort to understand the large deviations found by
\cite{abh66}, \kstar{} was added to the observing program at Ritter
Observatory.  From the first few spectra of \kstar{}, we noted that
the lines associated with the two components of \kstar{} (A \& B) were
broadened by different amounts.  As \cite{abh66} found the two
components of \kstar{} to have nearly identical masses and a
relatively short period, \kstar{} is expected to be tidally
interacting (\cite{tas92}).  This tidal interaction is predicted to
synchronize the orbital and rotational periods of the binary.
Thus, \kstar{} makes possible a sensitive test of synchronization
theory, as the two components have similar masses yet show different
rotational velocities.

  In $\S$2, we describe the observations.  The measurement of the
radial velocities, the computation of the $v\sin i$'s, and the
determination of the orbital parameters of \kstar{} are detailed in
$\S$3.  Section 4 contains a discussion of the results.

\section{Observations}

  The observations were carried out between May 1994 and July 1996.
The Ritter Observatory 1-m telescope was used in conjunction with an
\'echelle spectrograph connected to the Cassegrain focus of the
telescope by a fiber optic cable.  The spectrograph camera was
fabricated by Wright Instruments Ltd.\ and utilized a $1200 \times
800$ thick chip CCD with $22.5 \mu {\rm m}$ square pixels.  The CCD
was liquid-nitrogen cooled to an operating temperature of 140 K.  Data
acquisition was controlled by an IBM compatible personal computer
running software supplied by Wright Instruments, Ltd.

  The reduction of the data was done in the Interactive Data Language
(IDL) with a specialized program written for Ritter Observatory
spectra (\cite{gor96}) based on methods detailed in \cite{hal94}.  A
brief outline of the reduction process is as follows.  The average
bias and flat field were constructed on a pixel-by-pixel basis
allowing the removal of cosmic ray hits.  The average bias was
subtracted from the average flat field, object, and comparison frames.
The flat field was used to determine the order and background
templates.  The background template was used to remove the scattered
light from the flat field, objects, and comparisons after fitting a
polynomial to the inter-order background on a column by column basis.
Cosmic ray hits in the objects and comparisons were removed from
consideration by comparison with the average flat field.  The
normalized, smoothed, flat field profile was squared and multiplied by
the object profile, and the resulting function was summed in order to
obtain a profile-weighted extraction of the object spectrum.  The
wavelength calibration was accomplished by scaling all the comparison
lines to one super-order and fitting a polynomial to the result,
iteratively removing points until a preset standard deviation was
achieved.  Further information about Ritter Observatory (telescope,
instruments, reduction, and archive) can be found on the World Wide
Web (WWW) site {\em
http://www.physics.utoledo.edu/www/ritter/ritter.html}.

\begin{figure}[tbp]
\begin{center}
\plotfiddle{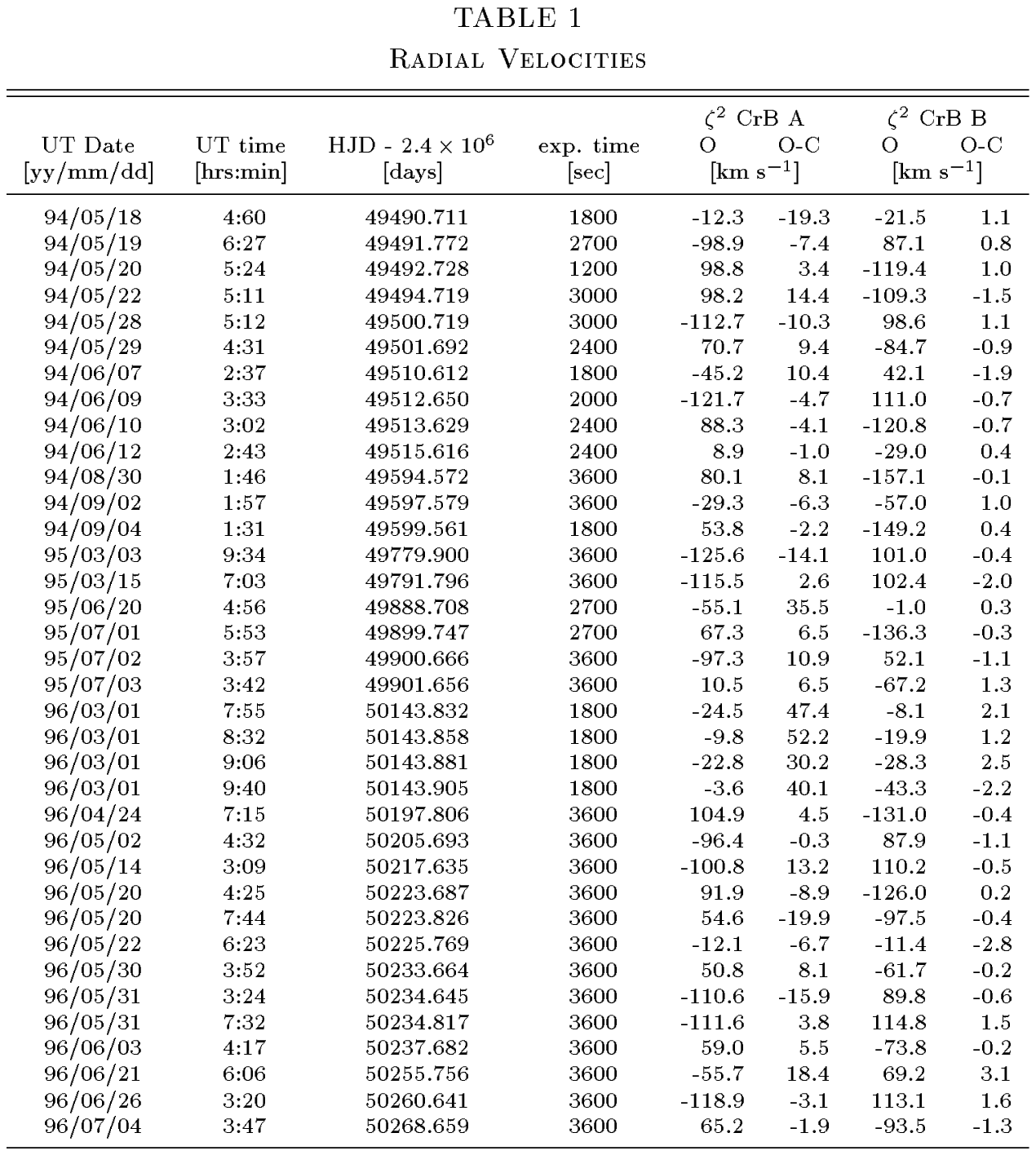}{5in}{0}{100}{100}{-300}{-200}
\end{center}
\end{figure}

  The observations of \kstar{} are tabulated in Table I, with the UT
date, UT time, HJD, and exposure time given in columns 1--4.  The
spectral coverage consisted of 9 disjoint orders between 5200 \AA\ and
6600 \AA\ with each order approximately 70 \AA\ wide.  Lines of
interest were H$\alpha$, He I 5876 \AA, and Si II 6347,6371 \AA.  The
slit width of the spectrograph was chosen to give a resolving power of
$R \approx 25,000$.  With this resolving power, the average
signal-to-noise ratio (S/N) of the observations of \kstar{} was about
100.  With this high S/N ratio, it was easy to distinguish the two
components of \kstar{} (narrow versus broad) by examining the
composite Si II 6371 \AA\ line.  As the Si II 6371 \AA\ line is a weak
line in B7 V stars, it was important to get high S/N observations.

\section{Analysis}

\subsection{Radial Velocities}

  The radial velocities were determined two ways, by fitting Gaussians
to individual lines and by cross-correlating the weak Si II 6371 \AA\
line.  The lines fit to Gaussians were H$\alpha$, Si II 6347, 6371
\AA, and He I 5876 \AA.  The results for each line were averaged
together after the rest wavelengths of H$\alpha$ and He I 5876 \AA\
were shifted until their radial velocities roughly matched the Si II
radial velocities.  This was done as both H$\alpha$ and He I 5876 \AA\
are multiplets with their rest wavelengths dependent on the different
strengths of the individual lines in the multiplet.

  The resulting average radial velocities determined from fitting
Gaussians were used in constructing templates for the
cross-correlation method.  The templates were constructed in a manner
similar to Cardelli \& Ebbets (1993).  Each observed spectrum was
assumed to include four sources - \kstar{} A (broad lines), \kstar{} B
(narrow lines), $\zeta^1$ CrB, and atmospheric lines.  On nights with
poor seeing, $\zeta^1$ CrB contributed to the spectrum due to the
large size of the fiber ($d = 5\arcsec$).  Atmospheric lines were
present in most of the spectra with strengths dependent on the water
content of the atmosphere.  A template was constructed for \kstar{} A,
\kstar{} B, the average $\zeta^1$ CrB spectrum, and an average
atmospheric spectrum.  Each template was constructed by iteratively
dividing each spectrum by the other three templates and then coadding
all the spectra in the rest frame of the template.  The spectra used
in constructing the templates were those in which the broad and narrow
lines were separated by over 100 km s$^{-1}$.  After 10 iterations, no
significant change in the templates was seen.  These templates were
used to cross-correlate the individual spectra and determine radial
velocities used in the rest of this paper.

  Table 1 lists the radial velocities determined from
cross-correlating the spectra with the \kstar{} A \& B templates.
Columns 5 \& 7 of Table 1 list the radial velocities for \kstar{} A \&
B, respectively.  The cross-correlation worked well for the narrow
lines of \kstar{} B, but for the broad lines of \kstar{} A it was
difficult choosing the right peak in the cross-correlation spectrum.
As a result, the error in \kstar{} A's radial velocity measurements
was $\approx 10$ km s$^{-1}$ and the error for \kstar{} B's radial
velocities was $\approx 1.5$ km s$^{-1}$ (see $\S$3.3).  While the
narrow line was not affected by pair blending in spectra when the two
lines were separated by less than 50 km s$^{-1}$, the broad line was
affected.  For these spectra the radial velocities measured by either
method for the broad line were systematically too close to the
velocity of the narrow line.

\begin{figure}[tbp]
\begin{center}
\plotone{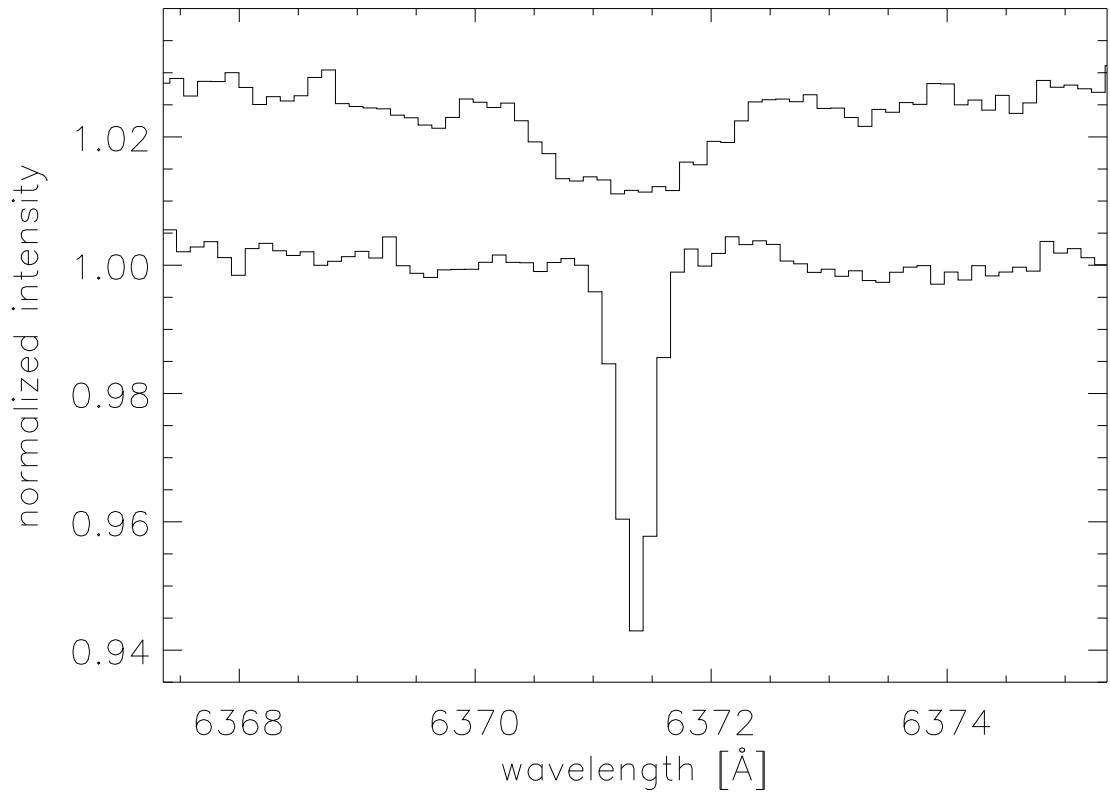}
\caption{The Si II 6371 \AA\ line in the templates for \kstar{} A
(top) and \kstar{} B (bottom) is plotted.  Note the difference between
the rotational broadenings of the line for \kstar{} A \&
B. \label{spectra:both}}
\end{center}
\end{figure}

\subsection{$V\sin i$}

  As the templates were constructed from many individual spectra, they
had a S/N greater than 500, a large improvement over the individual
spectra.  The region around the Si II 6371 \AA\ line in both templates
of \kstar{} A \& B is plotted in Fig.~{\ref{spectra:both}}.  As can be
seen from Fig.~{\ref{spectra:both}}, the line in \kstar{} A's spectrum
has the elliptical shape expected from fairly fast rotation, while the
line in \kstar{} B's spectrum has a Gaussian shape expected from
fairly slow rotation.  Using the templates, the half width at half
maximum (HWHM) was measured as $37 \pm 2$ km s$^{-1}$ for \kstar{} A
and $8 \pm 1$ km s$^{-1}$ for \kstar{} B.  These measured HWHM were
instrumentally debroadened using an instrumental HWHM of 5.5 km
s$^{-1}$ resulting in HWHM of $36 \pm 2$ km s$^{-1}$ and $6 \pm 1$ km
s$^{-1}$ for \kstar{} A \& B, respectively.  The $v\sin i$ values were
computed using the standard half-width method in which the unbroadened
line is assumed to be arbitrarily sharp but with a finite equivalent
width (\cite{col91}).  The value of the limb-darkening coefficient,
$\alpha$, can vary between 0.0 and 1.0 with the standard value being
0.6.  The values of $v\sin i$ corresponding to $\alpha =$ 0.0, 0.6, \&
1.0 were 41.7, 49.2, \& 51.0 km s$^{-1}$ for \kstar{} A and 6.7, 7.9,
\& 8.2 km s$^{-1}$ for \kstar{} B.  Using the standard method for
computing $v\sin i$ has been shown to be accurate to the 10\% level
for slowly to moderately rotating ($< 125$ km s$^{-1}$) early-type
stars (\cite{col95}).  Taking into account the error in the half width
and the range of possible values of $\alpha$, the values of $v\sin i$
were $46 \pm 7$ km s$^{-1}$ for \kstar{} A and $7.5 \pm 2$ km s$^{-1}$
for \kstar{} B.

\subsection{Orbit}

  Due to the difficulty in measuring the radial velocities of broad
lines of \kstar{} A, all but one of the orbital parameters for
\kstar{} were determined from the radial velocities of \kstar{} B.
The K amplitude of \kstar{} A was determined by assuming the \kstar{}
B's orbital fit and only fitting the K amplitude of the radial
velocities of \kstar{} A.  Orbital fits were done using the IDL
procedure CURVEFIT which was taken from Bevington (1969).  A more
specific program for computing spectroscopic orbits by \cite{wol67}
was also run producing similar results.

  Using all of \kstar{} B's radial velocities resulted in an orbit
with a period of 1.72 days, over a factor of 7 smaller than the
orbital period calculated by \cite{abh66}.  The residuals calculated
from the 1.72 day orbit were as large as 20 km s$^{-1}$, much higher
than could be attributed to measurement errors.  From measurements of
standard radial velocity stars taken on the same nights as the
\kstar{} observations, the radial velocity error for a narrow line was
about 0.2 km s$^{-1}$.  These large residuals could only be a result
of another period in the radial velocities of \kstar{} B.

  As B7 V stars are unlikely to have pulsations (\cite{wae91},
\cite{ste93}), a third star in \kstar{} was the most likely cause of
the residuals.  In order to possess a stable orbit, the third star
would need to have a significantly longer period than the inner binary
consisting of \kstar{} A \& B.  Therefore, an orbit for the inner
binary was fitted to observations closely spaced in time.  There were
two such data sets, HJD = 2449490.711 to 2449515.616 and HJD =
2450197.806 to 2450268.659.  The orbits fitted to these two data sets
had residuals on order of 1 km s$^{-1}$, a great improvement over the
previous fit using all the data.  The orbits determined from the two
data sets were essentially the same, except for a 1 km s$^{-1}$ shift
of their systemic velocities, which was within the systemic velocity
errors.  Combining both data sets resulted in an improved fit.  An
orbit for \kstar{} A was found by adopting all the orbital parameters
from \kstar{} B's fit, except for the K amplitude which was fit using
\kstar{} A's radial velocities.

  The residuals computed from observations where both \kstar{} A \&
B's radial velocities were at least 30 km s$^{-1}$ from the systemic
velocity were found to be correlated with a linear correlation
coefficient of 0.65.  Bevington (1969) gives a 0.1\% probability that
such a correlation coefficient involving 26 points would arise
randomly.  Thus, the residuals of \kstar{} A \& B have the same origin
and this provides concrete evidence that \kstar{} is actually a triple
system.

\begin{figure}[tbp]
\begin{center}
\plotone{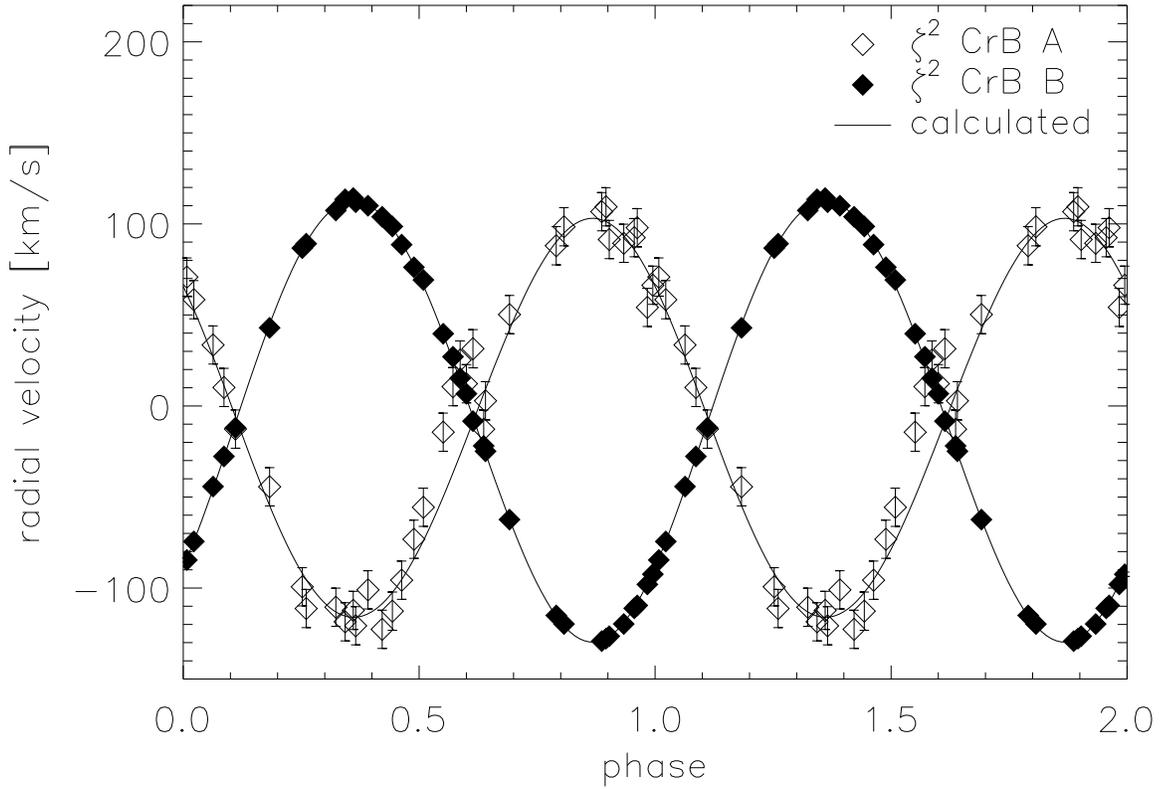}
\caption{The radial velocities for both \kstar{} A \& B are plotted
phased to the inner binary period, 1.72 days, given in Table 2.  The
fitted orbit to the outer binary has been subtracted from the radial
velocities plotted.  The solid lines are the fitted orbits using the
parameters listed in Table 2.  Note the large errors in the radial
velocities of \kstar{} A when it is near the systemic velocity of the
system.
\label{fig:inner_binary}} 
\end{center}
\end{figure}

\begin{figure}[tbp]
\begin{center}
\plotone{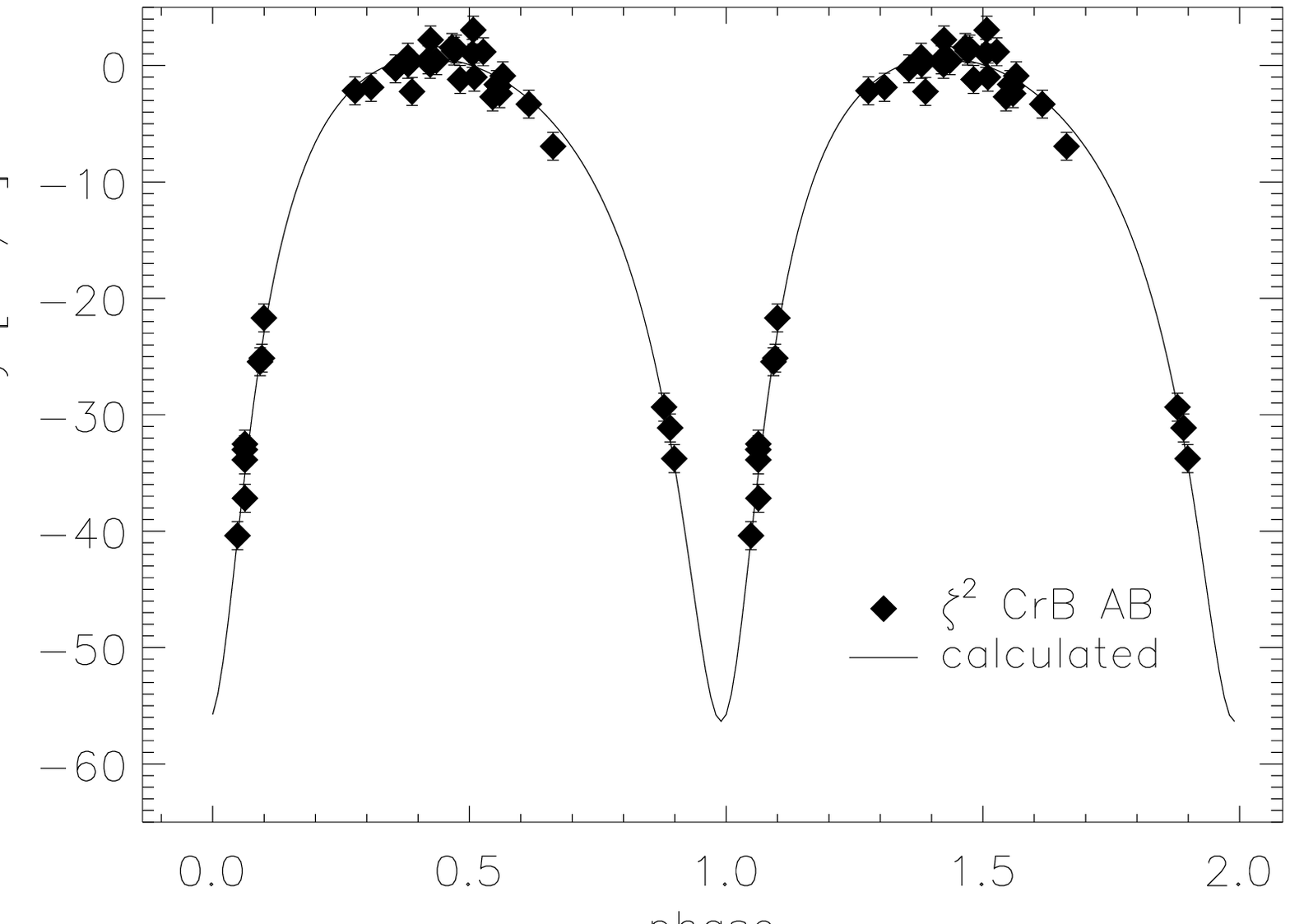}
\caption{The radial velocities for \kstar{} B are plotted phased to
the outer binary period, 251.5 days, given in Table 2.  The fitted
orbit to the inner binary has been subtracted from the radial
velocities plotted.  The solid line is the fitted orbit using the
parameters listed in Table 2. \label{fig:outer_binary}}
\end{center}
\end{figure}

  The outer binary, assumed to consist of the inner binary (\kstar{} A
\& B) and an unseen third star (\kstar{} C), was examined by looking
at the residuals of \kstar{} B's orbit.  As the residuals of \kstar{}
A \& B were correlated, changes in \kstar{} B's residuals were the
result of \kstar{} C's orbit around \kstar{} A \& B.  The phase
coverage was sufficient to derive an orbit, but this orbit was not
well determined due to the lack of observations between phases 0.95
and 1.05.  The orbital fits to both \kstar{} B and \kstar{} AB were
refined iteratively by subtracting the contribution from one orbitial
fit from \kstar{} B's radial velocities, fitting for the other orbit,
and repeating.  After the fifth iteration little change was seen in
the fitted orbital parameters.  The final orbital parameters for both
the inner binary and the outer binary are tabulated in Table 2.
Columns 6 \& 8 in Table 1 give the residuals (O-C) after subtracting
both inner and outer binary orbits.  These residuals were used to
estimate the errors in an individual radial velocity measurement
giving 10.5 km s$^{-1}$ for \kstar{} A and 1.2 km s$^{-1}$ for
\kstar{} B.  Figures~\ref{fig:inner_binary} \& \ref{fig:outer_binary}
plot the fitted orbits and radial velocities for both the inner and
outer binaries, respectively.  The orbital fits for the outer and
inner orbits have been removed from the radial velocities plotted in
Figures 2 \& 3, respectively.

\begin{table}[tbp]
\begin{center}
{\sc TABLE 2} \\
{\sc Orbit Parameters} \\[0.1in]
\begin{tabular}{cccc} \tableline\tableline
 & \kstar{} A & \kstar{} B & \kstar{} AB \\ \tableline
$V_o$ & \multicolumn{3}{c}{$-21.9 \pm 0.4$ km s$^{-1}$} \\ \tableline
$P$ [days] & \multicolumn{2}{c}{$1.72357 \pm 0.0001$} & $251.5 \pm 0.6$ \\
$T$ [days] & \multicolumn{2}{c}{$2450196.2793 \pm 0.0137$} & 
   $2449373.5 \pm 1.7$ \\
$e$ & \multicolumn{2}{c}{$0.013 \pm 0.002$} & $0.48 \pm 0.03$ \\ \tableline
$K$ [km s$^{-1}$] & $109.6 \pm 13.6$ & $121.2 \pm 0.3$ & $28.5 \pm 2.0$ \\
$\omega$ & $49\arcdeg \pm 3\arcdeg$ & $229\arcdeg \pm 3\arcdeg$ &
   $191\fdg 8 \pm 2\fdg 9$ \\
$a\sin i$ [R$_{\sun}$] & $3.73 \pm 0.46$ & $4.13 \pm 0.01$ & $124 \pm 7$ \\
$m\sin^3 i$ [M$_{\sun}$] & $1.155 \pm 0.142$ & $1.045 \pm 0.142$ & \nodata \\
\tableline\tableline
\end{tabular}
\end{center}
\label{table:param}
\end{table}

\subsection{Inclination}

  In non-eclipsing, unresolved binaries a common method to determine
the inclination is to assume one of the stars is synchronously
rotating.  This assumption implies the rotational and orbital periods
are equal and the rotational and orbital axes are parallel.  The
equatorial velocity of the star is computed from an assumed stellar
radius and then the inclination is computed from the star's measured
$v\sin i$.

  For \kstar{}, we applied this method assuming \kstar{} A (broad
lines) was synchronously rotating.  Assuming \kstar{} B (narrow lines)
synchronously rotates resulted in a equatorial velocity for \kstar{} A
larger than its breakup velocity.  The radius was assumed to be 2.38
R$_{\sun}$, but is only accurate to 50\% (\cite{and91}).  Using this
radius and the range in $v\sin i$ calculated in $\S$3.2, the resulting
range in $\sin i$ was 0.62--0.76.  Using these values of $\sin i$, the
range in masses was 4.83--2.64 M$_{\sun}$ for \kstar{} A and
4.37--2.39 M$_{\sun}$ for \kstar{} B.  For a B7 V star, Andersen
(1991) gives a mass of 4.13 M$_{\sun}$ with an accuracy of 15\%.
Thus, consistent values for \kstar{} A \& B radii and masses were
possible if the limb-darkening coefficient, $\alpha$, is fairly low.
Assuming an $\alpha$ of 0.0, gives $i = 38\arcdeg$.

  The mass function for the outer binary (\kstar{} AB \& C) was
computed to have a value of $0.42 \pm 0.11$ M$_{\sun}$.  Assuming that
all three components of \kstar{} are coplanar and have the range in
$\sin i$ calculated above, the mass function can be solved for
\kstar{} C's 
mass.  The mass range for \kstar{} C was computed to be 8.5--3.5
M$_{\sun}$.  Such a massive star would be visible in the spectrum of
\kstar{}.  Due to the lack of observations between the critical phases
0.95 and 1.05 of the outer binary orbit, the fitted K amplitude is
likely to be too large.  Reducing the K amplitude by a few km s$^{-1}$
would greatly reduce \kstar{} C's mass as the mass function is
proportional to $K^3$.

\section{Discussion}

  The significant results of this work were the discovery that
\kstar{} is a triple system, the inner binary has a much shorter
period than previously thought (\cite{abh66}), and \kstar{} B is
rotating asynchronously.

  The identification of \kstar{} as a triple system and the 1.72357
day inner binary period most likely explains the large residuals of
\cite{abh66}'s orbit.  \cite{abh66} calculated such a different
period, 12.5842 days, most likely for two reasons.  First, they only
calculated corrections to Plaskett's (1925) orbit.  Second, they only
measured H and He lines which, as multiplets, are intrinsically
broadened.  As a result, the H and He lines appear to have similar
widths making an identification of broad versus narrow difficult.
Only with high S/N spectra and a weak, intrinsically narrow line, such
as Si II 6371 \AA, were we able to distinguish consistently between
lines from \kstar{} A \& B.  In fact, \cite{abh66}'s data are
consistent with our fitted orbits with only a small number of their
points having wrong identifications.

  Using \cite{abh66}'s orbit, the lower limit on the masses of
\kstar{} A \& B ($m\sin^3 i$) were 9.9 M$_{\sun}$ and 9.4 M$_{\sun}$,
respectively.  \cite{abh66}'s lower limits on the masses are over a
factor of two greater normal mass of a B7 V star which is 4.13
M$_{\sun}$ (\cite{and91}).  Using our orbit,
the lower limit on the masses of \kstar{} A \& B are $1.155 \pm
0.142$ M$_{\sun}$ and $1.045 \pm 0.142$ M$_{\sun}$, respectively.
These lower limits are consistent with the mass of a B7 V star
clearing up the contradiction implied by \cite{abh66}'s work.

  The two components of the inner binary, \kstar{} A \& B, have equal
masses within the error bars, yet possess very different rotational
velocities.  From the work of Tassoul \& Tassoul (1992) and Claret et
al.\ (1995), the circularization and synchronization time-scales were
computed to be $10^6$--$10^7$ and $10^4$ years, respectively.  From
the above calculations, \kstar{} B should have synchronized its
rotation even before the inner binary circularized.  Obviously, some
other process is keeping \kstar{} B from synchronizing.  Claret \&
Gim\'enez (1995) were able to explain the asynchronous rotation in TZ
For as due to evolutionary effects.  Similarly, evolutionary effects
probably explain the asynchronous rotation of \kstar{} B.

  More observations of \kstar{} are needed, especially between phases
0.95 and 1.05 of the outer binary.  These observations would greatly
refine outer binary orbit, specifically its K amplitude and
eccentricity.

\acknowledgments

  This work was possible only with the help of the Ritter technician
Bob Burmeister and the crack Ritter observing team.  Team members
contributing observations to this effort were Jason Aufdenberg,
Michelle Beaver, Bruce Cantor, David Knauth, Alex Mak, Nancy Morrison,
Jens Petersohn, and both authors.  We are also thankful for many
helpful conversations with Nancy Morrison and Bernard Bopp.  Support
for observational research at Ritter Observatory is provided by NSF
grant AST-9024802 to B.\ W.\ Bopp and by The University of Toledo.
This research has made use of the Simbad database, operated at CDS,
Strasbourg, France.


\begin{thebibliography}{}

\bibitem[AS66]{abh66}
Abhyankar, K.\ D.\ \& Sarma, M.\ B.\ K.\ 1966, \mnras, 133, 437
[\cite{abh66}]

\bibitem[Andersen 1991]{and91}
Andersen, J.\ 1991, \aapr, 3, 91

\bibitem[Batten et al.\ 1989]{bat89}
Batten, A.\ H., Fletcher, J.\ M., \& MacCarthy, D.\ G.\ 1989,
Pub.\ Dominion Astrophys.\ Obs.\ Victoria, 17, 1

\bibitem[Bevington 1969]{bev69}
Bevington, P.\ R.\ 1969, Data Reduction and Error Analysis for the
Physical Sciences (New York: McGraw-Hill)

\bibitem[Cardelli \& Ebbets 1993]{car93}
Cardelli, J.\ A.\ \& Ebbets, D.\ C.\ 1993, in Calibrating Hubble
Space Telescope, ed.\ J.\ C.\ Blades \& S.\ J.\ Osmer (Baltimore:
STScI), 322

\bibitem[Claret \& Gim\'enez 1995]{cla95a}
Claret, A.\ \& Gim\'enez, A.\ 1995, \aap, 296, 180

\bibitem[Claret, Gim\'enez, \& Cunha 1995]{cla95b}
Claret, A., Gim\'enez, A., \& Cunha, N.C.S.\ 1995, \aap, 299, 724

\bibitem[Collins \& Cranmer 1991]{col91}
Collins, G.\ W.\ II \& Cranmer, S.\ R.\ 1991, \mnras, 53, 167

\bibitem[Collins \& Truax 1995]{col95}
Collins, G.\ W.\ II \& Truax, R.\ J.\ 1995, \apj, 439, 860

\bibitem[Gordon 1996]{gor96}
Gordon, K.\ D.\ 1996, Reduce95 Reduction Manual, in preparation

\bibitem[Hall et al.\ (1994)]{hal94}
Hall, J.\ C., Fulton, E.\ E., Huenemoerder, D.\ P., Welty, A.\ D., \&
Neff, J.\ E.\ 1994, \pasp, 106, 315 

\bibitem[Hoffleit 1982]{hof82}
Hoffleit D.\ 1982, Catalogue of Bright Stars, 4th Revised Ed.\ (New
Haven, CT: Yale Obs.)

\bibitem[Plaskett (1925)]{pla25}
Plaskett, J.\ S.\ 1925, Pub.\ Dominion Astrophys.\ Obs., 3, 179

\bibitem[Sterken \& Jerzykiewicz 1993]{ste93}
Sterken, C.\ \& Jerzykiewicz, M.\ 1993, SSRv, 62, 95

\bibitem[Tassoul \& Tassoul 1992]{tas92}
Tassoul, J.-L.\ \& Tossoul, M.\ 1992, \apj, 395, 259

\bibitem[Waelkens 1991]{wae91}
Waelkens, C.\ 1991, \aap, 246, 453

\bibitem[Wolfe, Horak, \& Storer (1967)]{wol67}
Wolfe, R.\ H., Horak, H.\ G., \& Storer, N.\ W.\ 1967, in Modern
Astrophysics: A Memorial to Otto Struve, ed.\ M.\ Hack (New York:
Gordon \& Breach)

\end{thebibliography}
\end{document}